\documentclass[%
 aip,
 amsmath,amssymb,
 reprint,%
]{revtex4-1}

\usepackage{graphicx}% Include figure files
\usepackage{dcolumn}% Align table columns on decimal point
\usepackage{bm}% bold math
\usepackage[utf8]{inputenc}
\usepackage[T1]{fontenc}
\usepackage{mathptmx}

\begin{document}

\preprint{AIP/123-QED}

\title[]{Silicon-nitride nanosensors toward room temperature quantum optomechanics}

\author{Enrico Serra}
 %\altaffiliation[Also at ]{Institute of Materials for Electronics and Magnetism, Nanoscience-Trento-FBK Division, 38123 Povo, Trento, Italy}
\email{E.Serra@tudelft.nl}
\affiliation{Istituto Nazionale di Fisica Nucleare (INFN), Trento Institute for Fundamental Physics and Application, 38123 Povo, Trento, Italy}
\affiliation{Institute of Materials for Electronics and Magnetism, Nanoscience-Trento-FBK Division, 38123 Povo, Trento, Italy}
\affiliation{Department of Microelectronics, Delft University of Technology, Feldmannweg 17, 2628 CT Delft, The Netherlands}
\author{Antonio Borrielli}
%\homepage{http://www.Second.institution.edu/~Charlie.Author.}
\affiliation{Institute of Materials for Electronics and Magnetism, Nanoscience-Trento-FBK Division, 38123 Povo, Trento, Italy}
\affiliation{Istituto Nazionale di Fisica Nucleare (INFN), Trento Institute for Fundamental Physics and Application, 38123 Povo, Trento, Italy}
\author{Francesco Marin}
\affiliation{European Laboratory for Non-Linear Spectroscopy (LENS), Via Carrara 1, I-50019 Sesto Fiorentino (FI), Italy }
\affiliation{CNR-INO, L.go Enrico Fermi 6, I-50125 Firenze, Italy}
\affiliation{INFN, Sezione di Firenze}
\affiliation{Dipartimento di Fisica e Astronomia, Universit\`a di Firenze, Via Sansone 1, I-50019 Sesto Fiorentino (FI), Italy}
\author{Francesco Marino}
\affiliation{CNR-INO, L.go Enrico Fermi 6, I-50125 Firenze, Italy}
\affiliation{INFN, Sezione di Firenze}
\author{Nicola Malossi}
\affiliation{Physics Division, School of Science and Technology, University of Camerino, I-62032 Camerino (MC), Italy} 
\affiliation{INFN, Sezione di Perugia, via A. Pascoli, I-06123 Perugia, Italy}
\author{Bruno Morana}
\affiliation{Department of Microelectronics, Delft University of Technology, Feldmannweg 17, 2628 CT Delft, The Netherlands}
\author{Paolo Piergentili}
\affiliation{Physics Division, School of Science and Technology, University of Camerino, I-62032 Camerino (MC), Italy} 
\affiliation{INFN, Sezione di Perugia, via A. Pascoli, I-06123 Perugia, Italy}
\author{Giovanni Andrea Prodi}
\affiliation{Dipartimento di Fisica, Universit\`a di Trento, I-38123 Povo, Trento, Italy}
\author{Lina Sarro}
\affiliation{Department of Microelectronics, Delft University of Technology, Feldmannweg 17, 2628 CT Delft, The Netherlands}
\author{Paolo Vezio}
\affiliation{CNR-INO, L.go Enrico Fermi 6, I-50125 Firenze, Italy}
\affiliation{INFN, Sezione di Firenze}
\author{David Vitali}
\affiliation{Physics Division, School of Science and Technology, University of Camerino, I-62032 Camerino (MC), Italy} 
\affiliation{INFN, Sezione di Perugia, via A. Pascoli, I-06123 Perugia, Italy}
\author{Michele Bonaldi}
\affiliation{Istituto Nazionale di Fisica Nucleare (INFN), Trento Institute for Fundamental Physics and Application, 38123 Povo, Trento, Italy}
\affiliation{Institute of Materials for Electronics and Magnetism, Nanoscience-Trento-FBK Division, 38123 Povo, Trento, Italy}

\date{\today}% It is always \today, today,

\begin{abstract}
Observation of quantum phenomena in cryogenic, optically cooled mechanical resonators has been recently achieved by a few experiments based on cavity optomechanics. A well-established experimental platform is based on a thin film stoichiometric ($ Si_3 N_4 $) nanomembrane embedded in a Fabry-Perot cavity, where the coupling with the light field is provided by the radiation pressure of the light impinging on the membrane surface. Two crucial parameters have to be optimized to ensure that these systems work at the quantum level: the cooperativity $ C$ describing the optomechanical coupling and the product $ Q \times \nu$ (quality factor - resonance frequency) related to the decoherence rate. A significant increase of the latter can be obtained with high aspect-ratio membrane resonators where uniform stress dilutes the mechanical dissipation. Furthermore, ultra-high $ Q \times \nu$ can be reached by drastically reducing the edge dissipation via clamp-tapering and/or by soft-clamping, virtually a clamp-free resonator configuration. In this work, we investigate, theoretically and experimentally, the edge loss mechanisms comparing two state-of-the-art resonators built by standard micro/nanofabrication techniques. The corresponding results would provide meaningful guidelines for designing new ultra-coherent resonating devices.
\end{abstract}

\maketitle

%\begin{quotation}
%The ``lead paragraph'' is encapsulated with the \LaTeX\ 
%\verb+quotation+ environment and is formatted as a single paragraph before the first section heading. 
%(The \verb+quotation+ environment reverts to its usual meaning after the first sectioning command.) 
%Note that numbered references are allowed in the lead paragraph.
%%
%The lead paragraph will only be found in an article being prepared for the journal \textit{Chaos}.
%\end{quotation}

\section{\label{sec:level1} INTRODUCTION}

A significant boost in the performance of sensors and metrological protocols can come from the use of quantum optomechanical sensors. 
In fact, quantum-enhanced metrology allows measurements with a precision that surpasses any classical limitation,
enabling a generation of sensors of unprecedented sensitivity. 
Optomechanical resonators made possible the measurement of continuous force and displacement below the standard quantum limit 
with SiN membranes,\cite{Mason19} sensing forces at the attonewton level  with phononic crystal silicon nitride nanobeams,\cite{Ghadimi18} the quantum non demolition measurement of optical field fluctuations with a highly reflective silicon 
resonator.\cite{Pontin18} The domain of application of quantum-enhanced devices ranges from  fundamental physics experiments, 
as the search for quantum gravity effects or the detection of weak stochastic forces,\cite{Bawaj15,Pontin14} to quantum 
technology applications. In fact, adding an electrical degree of freedom (like an electrode) to the mechanical device, 
one can realize bidirectional frequency electro-opto-mechanical modulators to convert optical signals,  
or  ultrasensitive trasducers for a shot-noise limited nuclear magnetic resonance (NMR).\cite{Moaddel18} 
These  kind of hybrid-devices are gaining momentum in foreseen quantum networks due to their peculiar property of 
maintaining the quantum coherence  for very long times.\cite{Andrews14}  

A wide variety of mechanical oscillators can be designed and coupled with a light field, like silicon optomechanical crystals, 
silica microtoroids, silicon nitride nanobeams or nanomembranes, with masses ranging from $ \mu g$ to pg and frequencies from kHz 
to GHz.\cite{Aspelmeyer14} 
In general, quantum properties of all optomechanical system are hidden and/or destroyed by thermal noise. For this reason, all quantum 
effects observed to date have been obtained on resonators placed in a cryogenic environment. Unfortunately this requirement is a major 
obstacle to the realization of exploitable sensors, therefore a recent branch of research
and few recent experiments on levitated nanospheres,\cite{Tebbenjohanns20,Delic20,Magrini21,Rafagni21} are progressing toward 
a new generation of optomechanical systems, able to maintain a quantum behavior even at room temperature.
Actually, to facilitate room temperature optomechanics, two major parameters should be improved: the optomechanical cooperativity $ C$, 
proportional to the optomechanical interaction strength, and the $ Q \times \nu$ product, 
inversely proportional to the coupling between the resonator and the  surrounding thermal bath. When the cooperativity $ C$
is larger than the number of quanta in the mechanical resonator $ n_{th} \cong k_B T /\hbar \nu$, the phonon-photon transfer 
is faster than the decoherence rate due phonon leakage from the resonator to the thermal bath. Cooperativity is defined as the ratio 
between the square of the coherent coupling rate g and the product between the optical ($ \kappa$) and mechanical ($ \gamma$) 
dissipation rates: $ C =4 g^2/\kappa \gamma$ where $ g=g_0 \sqrt{n_c}$, with  $ n_c$ the intracavity photon number,
 $ g_0$ the single-photon coupling rate. Hence, improving cooperativity means reducing losses and improving the optomechanical interaction. 
In turn, reducing losses improves the number of coherent oscillations the resonator can undergo before one environmental phonon enters the 
system, which is just the product $ Q \times \nu$. If we compare this figure with the thermal decoherence rate  $k_B T/(Q\hbar)$,
 we easily find  the condition for neglecting thermal decoherence over one mechanical period, that is  $ Q \times \nu >$  6.2 THz at 
 room temperature.

Usually the resonators is embedded into a high-finesse optical Fabry-Perot cavity, where interaction between photons and phonons takes place. 
The cavity is externally pumped with a monochromatic laser source (for example a Nd:YAG IR laser) and the radiation pressure exerted by
 the optical field on the vibrating device  generates intensity cavity field fluctuations containing information about its displacement. 
This technique has been used to probe a wide variety of physical phenomena.\cite{Aspelmeyer14} In the classical domain we have studied 
parametric squeezing of mechanical motion induced by optical field modulation,\cite{Pontin14b} 
while in the quantum domain we have harnessed the optomechanical interaction to make a quantum non-demolition measurements of optical 
fields,\cite{Pontin18} and more recently we explored the border between classical and quantum physics by observing the quantum signature 
of a squeezed mechanical oscillator.\cite{Chowdhury20} 

In this work we focus on resonators based on  stoichiometric $ Si_3N_4$ membranes, commonly used in optomechanical setups. Here 
to comply with the requirement  $ Q \times \nu >$  6.2 THz,  needed for room temperature cavity optomechanics,
the membrane is tensioned to dilute the mechanical dissipation, in a scheme first proposed in the mirror suspensions of 
gravitational wave antennae.\cite{Saulson90} Thanks to this effect, nanobeams or nanomembranes made of thin nitride amorphous 
films, with intrisic Q-factor of 4000, can reach ultra-high Q-factors  of the order of $10^8$ to $10^9$, 
if produced with an internal stress of about 1 GPa.\cite{Unter10,Schmid11} However the diluition of the dissipation is a design strategy that can be 
exploited also in other materials and can be complemented by other strain engineering approaches. For instance, in  III-V semiconductors 
($ GaAs$, $ GaNAs$, and $ In_{1-x}Ga_xP$),  strain tunability is obtained by lattice mismatch between the films epitaxially grown and the 
substrate  and also in this case the resulting  tensile stress can exceed 1 GPa.\cite{Weig18}  

In the following we  compare the performances of two state-of-the-art designs for membrane resonators, considering the stress engineering 
framework, the fabrication issues and their ultimate performances in view of room temperature optomechanics applications. 
We set out design rules that we plan to use in developing future ultra-high coherence devices.

%---------------------------------------------
\section{\label{sec:damping} DAMPING ANALYSIS}
%---------------------------------------------
The total mechanical dissipation can be obtained as the incoherent sum of extrinsic (for instance gas damping and clamping losses), 
and internal contributions. Internal losses can be classified into the following categories: intrinsic dissipation originated from 
the delay between strain and stress in the material, thermoelastic loss with local heat generation and conduction, and Akhiezer 
damping due to coupling between the strain field and the phonon modes. In  devices based on $Si_3N_4$ membranes the intrinsic 
dissipation determines the ultimate sensitivity of the device, as termoelastic and Akhiezer 
losses are at least one order of magnitude lower then the instrisic losses in case of silicon nitride.\cite{Ghaffari12}  
Clamping losses are also negligible in the considered devices, thanks to the use of on-chip seismic filtering stage  or
phononic bandgap crystal  effectively isolating the membrane from its frame.\cite{Borrielli16,Tsaturyan17}
%-------------
\begin{figure}
\resizebox{0.45\textwidth}{!}{%
  \includegraphics{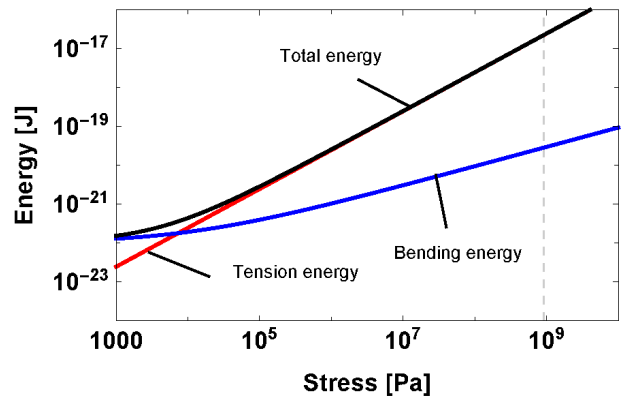}}
\caption{Energy diagram for the (0,1) mode of a circular-shaped nanomembrane resonator with dimensions 
$ D=1480 \ \mu$m, thickness $ h=100 \ n$m showing the contribution of the tensile, the bending energy, 
and the total energies by varing the internal stress $\sigma_0$.}
\label{fig:fig1}       
\end{figure}
%-------------
%-----------------------------------------------------------------------
\subsection{\label{sec:damping1} Dissipation dilution by uniform stress}
%-----------------------------------------------------------------------
A static strain $\epsilon_0$ originates in a $ Si_3N_4$ film from the mismatch between the linear expansion coefficient 
of the nitride layer and the substrate during the cooling phase that follows the Low Pressure Chemical Vapor Deposition (LPCVD) 
process. As a consequence, a biaxial state of stress $\sigma_0$ builds up in the film and reduces the intrinsic mechanical dissipation. 
After the first observations by Southworth,\cite{Southworth09} this effect has been theoretically investigated for 1D or 2D  
micro- and nano-mechanical resonators (strings or membranes) showing that the stress "dilutes" intrinsic 
dissipation, thanks to the increase of the oscillator resonant frequency while bending losses remain essentially constant.\cite{Schmid11, Yu2012} 
A more recent approach   made clear that the Q-factor increase can be evaluated  as the ratio between  
the total elastic energy due to internal stress $\sigma_0$ and the bending energy:\cite{Fedorov19}
%-------------
\begin{equation}
\label{eq:Q_Qint_W_t_W_bendtot}
\frac{Q}{Q_{int}}=   \frac{W_{tens}}{W_{{bend-edge} }+W_{bend-internal}}
\end{equation}
%-------------
where $ Q_{int}$ is the intrinsic Q-factor (when $\sigma_0=0 $) and in the bending energy, we have separated the contribution 
from the edge to that of the internal region of the resonator. Fig.~\ref{fig:fig1} shows the energy balance for the fundamental 
mode of a 2-D membrane with dimension $ D=1480 \ \mu m$ and thickness $ h=100 \ $nm as a function of the tensile stress. 
As it is possible  to fabricate films with a stress level in excess of 1 GPa, it is easy to produce a device dominated by 
tensile energy.
%--------------
\begin{figure}
\resizebox{0.45\textwidth}{!}{%
  \includegraphics{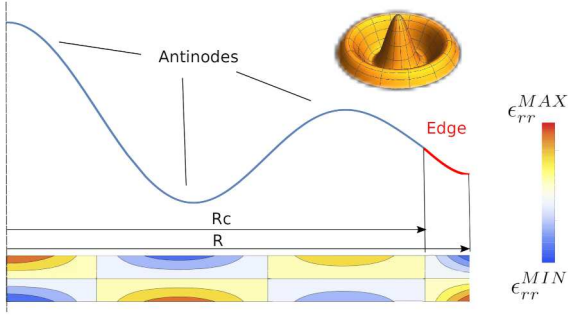}}
\caption{Cross-section of a circular mebrane showing the radial bending strain ($ \epsilon_{rr}$) countour plot 
along thickness in case of an axysimmetric mode (0,3). Antinodes strain 
(distributed loss) in the region $ (0,R_c)$ and edge strain (edge loss) in the 
region $ (R_c,R)$ (not in scale) are shown.}
\label{fig:fig2}       
\end{figure}
%--------------
%------------------------------
\begin{figure}
\resizebox{0.45\textwidth}{!}{%
\includegraphics{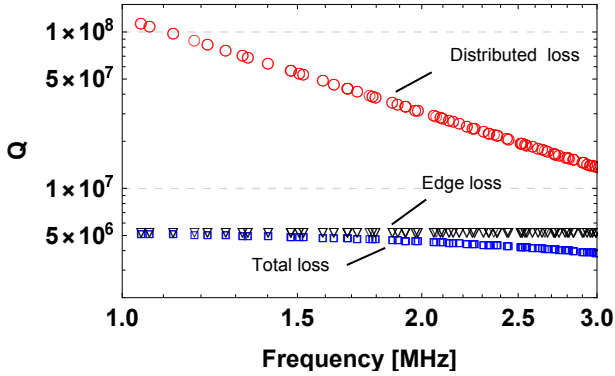}}
\caption{Expected Q-factor for a  circular-shaped membranes (dimensions $ D=1480 \ \mu m$, thickness 
$ h=100 \ nm$. $ \sigma_0=0.915 \ GPa$) when edge (black-triangles), distributed (red-circles) or total  
loss (blue-squares) contributions are considered.
}
\label{fig:fig3} 
\end{figure}
%-----------------------------------------------------------------
\subsection{\label{sec:damping2} Edge and distributed dissipation}
%-----------------------------------------------------------------
As pointed out by Schmid and Yu,\cite{Schmid11,Yu2012} intrinsic dissipation can be separated in  the sum of two 
fundamental contributions: one due to the bending at the clamping edge of the membrane and the other one coming from the 
internal area, mainly at the antinodes where local bending is higher as reported in Fig. \ref{fig:fig2} for a circular membrane.
Edge dissipation is frequency indipendent while the internal contribution increases with modal index as it is shown in Fig. \ref{fig:fig3}. 
Analytical approaches for evaluating the total intrinsic loss were first developed for  1-D doubly clamped beams and then extended to 
2-D nanomembranes,\cite{Yu2012} while in  a general approach for numerically estimating intrisic dissipation 
of 3-D structures is discussed.\cite{Fedorov19} In the anelastic loss model the stress $\sigma$ is delayed with respect to the strain 
$\epsilon$: $ \sigma=(Y+i \phi Y) \epsilon$, where the loss angle is related to the Q-factor as $ \phi=Q^{-1}$, therefore  
the dissipation is determined by the mean curvature. To clarify the role of edge dissipation in the total loss budget, we focus 
on the curvature in case of 1-D tensioned doubly-clamped nanostring of length L, thickness k and width w: 
%---------------
\begin{equation}
\label{eq:u_n}
Y I_z \frac{d^4u_n}{dx^4}- \sigma_0 L w \frac{d^2 u_n}{d x^2}-\rho \omega_n^2 u_n=0\\
\end{equation}
%---------------
%---------------
where $ Y I_z=\frac{Y h^3 w}{12}$ is the flexural rigidity, $ u_n$ is the modal shape function and $\sigma_0$ the internal stress. 
At the clamping point the inertial term  $ \rho\, \omega_n^2 u_n $ is negligible, and the mean curvature at the edge can be 
approximated as: 
%---------------
\begin{equation}
\label{eq:u_nCl}
\frac{d^2 u_{cl,n}(x)}{dx^2}=\left(\frac{du_n}{dx}\right)_{x=+0} \frac{1}{\lambda L}\exp{(-\frac{x}{\lambda L})}
\end{equation}
%---------------
where $ \lambda=\sqrt{1/12 \epsilon_0} \ h/L$ is known as dilution factor, essentially the  ratio between bending and elongation energies. 
From Eq. (\ref{eq:u_nCl}) the shape function at the clamping edge experiences a sharp exponential bending.
The derivative $ du_n/dx(0+)$ cannot be zero, because far from the clamping points the bending term can be neglected and the solution 
of Eq. (\ref{eq:u_n}) has a sinusoidal shape, that has to be connected with the solution at the clamping $ u_{cl,n}$ containing the exponential term.  
In high-strain limit we have $ \lambda \ll 1$ both in string and membranes and the critical region close to the clamping edges can be estimated as:
$ L_{cl}=\sqrt{Y /(12 \sigma_0)} h  \approx 5 h= 500 \ $nm for  h=100 nm. Therefore, the edge loss involves only a small region where major 
dissipation takes place. The procedure described above can be extented to 3-D square membrane of side $ L$ by
solving the general equation for vibrational plates written in terms of adimensional coordinates $ (\eta,\chi)$: 
%---------------
\begin{equation}
\label{eq:GE_Vibs}
\lambda^2 \nabla^4_{\eta,\chi} w_{nm}-\nabla^2_{\eta,\chi} w_{nm}- \Omega^2 w_{nm} =0
\end{equation}
%---------------
where $w_{nm}$ represents the displacement in the direction orthogonal to the membrane's plane, $ \Omega$ the normalized frequency 
and $ \lambda^2=D/(\sigma_0 \,h L^2)$ is still the dilution parameter, now expressed in terms of the flexural rigidity of the plate 
$D$. From Eq. (\ref{eq:GE_Vibs}) the Q-factor has  been derived  for a square-shaped membrane 
as the sum of the edge and distributed contributions:\cite{Yu2012}   
%---------------
\begin{equation}
\label{eq:Q_Qint_M_B}
\frac{Q}{Q_{int}}=(2 \lambda+ (n^2+m^2) \pi^2 \lambda^2)^{-1} 
\end{equation}
%---------------
where $ Q_{int}$ is the intrinsic Q-factor (without dilution effect), $ n,m$ the modal indexes. 
In the following of the paper we analyze and discuss the interplay between these two dissipation mechanisms
for the two devices presented in this work. 
%----------------------------------------------------------------------
\subsection{\label{sec:damping3} Circular membranes and modal expansion}
%----------------------------------------------------------------------
In case of a circular membrane of radius $R$, the solutions of Eq. \ref{eq:GE_Vibs} are the Bessel functions $J_m$
%---------------
\begin{equation}
 w_{mn}(r,\theta)=k_{mn} J_m(\alpha_{mn} \frac{r}{R})\cos(m \theta)
\end{equation}
Where $(r, \theta)$ is  the cylindrical coordinate system centered in the membrane with $\bf{i}_z$ aligned along the membrane's 
out-of-plane axis. $\alpha_{mn}$ is the $n$-th root of the Bessel polynomial of order $m$.
%---------------
The constant $k_{mn}$ is fixed by the  normalization requirement:
%---------------
\begin{equation}
h \int_S dS \, \rho w_{mn}^2(r,\theta) =M
\label{normalization}
\end{equation}
as
%---------------
\begin{equation}
k_{mn}=\sqrt{\frac{\pi R^2}{\int_S dS J_m^2(\alpha_{mn} \frac{r}{R})\cos^2(m \theta)}}.
\end{equation}
%---------------
where $h$ is the membrane's thickness and $M$ its mass.  
In the context of the modal expansion theory,\cite{meirovitch} spatial variables can be separated from the temporal ones. 
Therefore the displacement of the membrane ${ u}({ r},\theta, t)$ can be  written as a superposition of normal modes with time 
dependent coefficients
%---------------
\begin{equation}
{u}({r},\theta, t)=\sum_{mn} { a}_{mn}(t) \,{ w}_{mn}({ r},\theta)
\label{eq:espansione}
\end{equation}
%--------------- 
The membrane is then sampled on its surface by a readout with output $X$
\begin{equation}
X(t)=\int_S dS\; {P}_s({r},\theta ) \,{u}({r},\theta,t) 
\label{eq:XresponseV}
\end{equation}
where $ {P}_s$  is the weight function of the readout, normalized such that $\int_S dS\; {P}_s({r},\theta )=1$ and depending on the  measurement technology adopted. For instance, an optical 
readout based on laser beam is sensitive to the position of the membrane's surface, averaged over the gaussian beam profile 
with a waist $ w_0$: 
%---------------
\begin{equation}
 P_s(r,\theta)=\frac{2}{\pi w_0^2} e^{-2 r^2/w_0^2}  \equiv\frac{2}{\pi w_0^2} e^{-2 (x^2+y^2)/w_0^2} 
\label{eq:G}
\end{equation}  
%---------------
The normalization choice Eq. (\ref{normalization}) allows the convenient way of describing the motion of each mode ${ w}_{mn}$ as 
a harmonic oscillation with effective mass:
\begin{equation}
  m_{mn}^{\rm eff}=\frac{M}{(\int_S dS\, {P}_s(r,\,\theta) \;{w}_{mn}(r,\,\theta))^2}   
\label{effmass}
\end{equation}
From this equation it is easy to see that the lowest value of the effective mass is obtained when the readout 
samples a region of maximum oscillation amplitude (anti-node), while ${ m}_{mn}^{\rm eff}$ increases if the readout 
samples a nodal region. For the $(0,n)$ axisymmetrical modes read by a laser beam centered on the membrane axis, 
the effective mass is:
%---------------
\begin{equation}
  m_{0n}^{\rm eff}=\frac{M J_1^2(\alpha_{0n})}{2 (\pi \int_0^{R} P(r) J_0(\alpha_{0n}\frac{r}{R}) r dr)^2}
\label{modalMassCircle}
\end{equation}
%---------------
where $ M=\pi R^2 h \rho$ is the membrane physical mass, $ J_0, J_1$ the two lowest order Bessel functions 
and $ \alpha_{0n}$ the $ n-th$ root of $ J_0$. 
We remind that, in the pair of indexes $(m,n)$, the index $m$ is the number of radial nodal lines and $n$ is the number of circumferential nodal lines. 
In the lowest frequency drum mode (0,1),  the circumferential node is the fixed edge.
If the waist of the laser is much smaller than the membrane's radius, the first values of $ m_{0n}^{\rm eff}/M$ are 0.269, 0.116, 0.074, 0.054. 
We note that the effective mass of the drum mode (0,1) is about 1/3 of the total membrane's mass.
%------------------------------------------------------------------
\subsection{\label{sec:damping4} Soft-clamped rectangular membranes}
\begin{figure}
\resizebox{0.45\textwidth}{!}{%
  \includegraphics{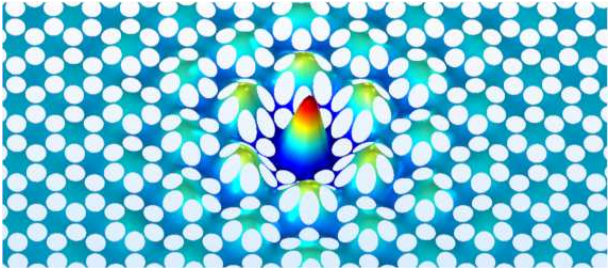}}
\caption{Modal shape function of a bandgap mode A evaluated by a 3-D finite element simulation.}
\label{fig:fig4}      
\end{figure}

\begin{figure}
\resizebox{0.45\textwidth}{!}{%
  \includegraphics{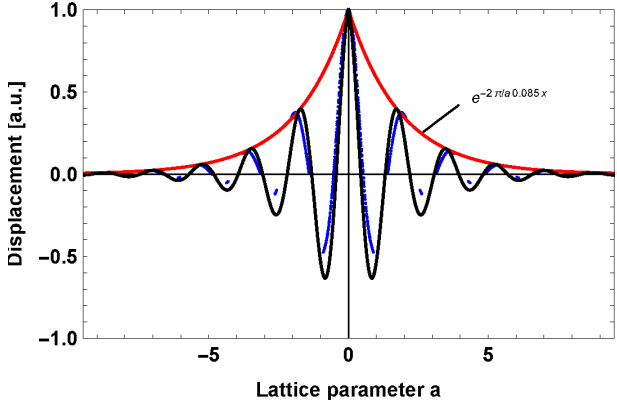}}
\caption{The bandgap mode A cross-section along $ x$ axis. Blue curve is derived by FEM modal analysis, the red curve is an exponential function as a guide to the eye and the black is the sinusoidal decaying approximation, showing the evanescent decay.}
\label{fig:fig5}      
\end{figure}

%-------------------------------------------------------------------
An effective method to reduce edge loss is to localize the vibrational modes away from the clamping region by soft-clamping.
In this way we achieve a clamping-free device where edge losses can be neglected. Soft-clamping is a novel design strategy based on phononic crystals, proposed by Tsaturyan in case of a square-shaped membrane,\cite{Tsaturyan17} and then applied  to nanostrings.\cite{Fedorov19,Kippenberg19} 
The bandgap of the phononic crystal is matched to the resonant frequency of the so called "defect modes" in the central 
part  of the phononic structure (Fig.~\ref{fig:fig3} and Fig.~\ref{fig:fig4}).
 A defect mode evanescently couple to the frame therefore eliminating the extra curvature at the edge imposed by the boundary conditions. 
Fig.~\ref{fig:fig5} shows that the effectiveness  of this soft-clamping depends on the number of phononic cells on the device:
 for instance with nine cells  the displacement at the edge  is reduced at least a factor of 100 w.r.t. the centre of the defect region. 
The Q-factor is limited only by the distributed losses when the edge contribution of the bending energy is neglected in Eq. (\ref{eq:Q_Qint_W_t_W_bendtot}).  As a consequence, Q-factors in the bandgap region reach values in excess of $10^8$.
Outside the bandgap the Q-factor of modes drop and is still limited by edge losses. 

For a 1D string resonator of thickness $ h$, Eq. (\ref{eq:Q_Qint_W_t_W_bendtot}) for the mode $ u_n$ is: 
%------------
\begin{eqnarray}
\label{eq:Q_Qint_soft-clamping}
 Q&=&  Q_{int}  \frac{W_{tens}}{W_{{bend-internal}}} \nonumber \\
 &=&   Q_{int}  \frac{\int_{0}^{L}\sigma(x) A(x) (\frac{du_n(x)}{dx})^2 dx}{\int_0^{L} Y I_z(x) (\frac{d^2 u_n(x)}{dx^2})^2 dx} \nonumber \\
 &=&  \eta \, \beta \, \frac{\sigma_0}{Y} \frac{a^2}{h}
\end{eqnarray}
%------------
where $\sigma_0$ is the average stress, $ A(x)=h(x)w$ the cross-section area, $ a$ the lattice parameter, $ L$  the string length, $ \eta$ a pre-factor related to the mode shape and $\beta$ is the average surface loss parameter. 
%-----------------------------------------
%-----------------------------------------
%-----------------------------------------
We point out that  in this equation the intrinsic dissipation contribution is mainly due to surface losses $ Q_{int}^{-1}=\frac{1}{\beta h}$.
In fact surface loss is one of the  damping mechanisms in micro- and nano-mechanical nitride resonators and its contribution dominates 
when the thickness $ h$ is less than  $100$ nm.\cite{Villanueva14}

For a nanomembrane the analogous of the Eq. (\ref{eq:Q_Qint_soft-clamping}) becomes: 
%------------
\begin{eqnarray}
\label{eq:Q_Qint_soft-clamping2D}
 &&  {Q}= {Q_{int}} \frac{W_{tens}}{W_{{bend-internal}}}= \nonumber\\
 &=& {Q_{int}} \frac{h\int_{-L_1/2}^{L_1/2} \int_{-L_2/2}^{L_2/2} \sigma(x,y) ((\frac{\partial w_{BG}(x,y)}{\partial x})^2+(\frac{\partial w_{BG}(x,y)}{\partial y})^2) dx dy}{D_z \int_{-L_1/2}^{L_1/2}\int_{-L_2/2}^{L_2/2}   (\frac{\partial^2 w_{BG}(x,y)}{\partial x^2}+\frac{\partial^2 w_{BG}(x,y)}{\partial y^2})^2 dx dy} \nonumber \\
&=&  f(a) \times \frac{\sigma_0}{Y \ h^2}\times\frac{1}{\phi + \frac{1}{\beta h}}
\end{eqnarray}
%------------
where $ w_{BG}$ is the modal shape function of a defect mode, $ f(a)$ is numerically evaluated from the mode shape as a function of the lattice constant $ a$ and of the membrane size. 
In this case bulk  loss parameter $\phi$ and surface loss parameter $\beta$ concur to determine the overall dissipation of the mode. 
For a membrane with dimensions $  L_1=19\,a$, $  L_2=19.5\,a$, the shape of the bandgap mode A is:
%------------
\begin{eqnarray}
\label{eq:approxmodeA_2D}
 &&  w_A(x,y)=  k_A e^{(-2 \pi/a \ 0.085 \ (|x|+|y|))} \times \nonumber\\
&& \times \cos{(2 \pi /a \ 0.57 \ |x|)}\cos{(2 \pi /a \ 0.57 \ |y|)} 
\end{eqnarray}
where $k_A$ is detemined by the normalization requirement Eq. \ref{normalization}. This shape function is compared in Fig.~\ref{fig:fig4} with a FEM evaluation.
%------------ 
The effective modal mass for mode A can be estimated by numerically evaluating the eigenfunction by finite elements or  from the modal shape $ w_A(x,y)$ as:
%---------------
\begin{equation}
 m^{\rm eff}_A = \frac{M}{(\int_{-L_1/2}^{L_1/2} \int_{-L_2/2}^{L_2/2} P(x,y)\, w_i(x,y) \,dx dy)^2}
\label{modalMassA}
\end{equation}
%--------------

%-----------------------------------------------------------
\section{Chip-scale optomechanical $ Si_3N_4$ oscillators}
%-----------------------------------------------------------
In the following we describe the fabrication of a round-shaped nanomembrane with an on-chip isolation stage 
\cite{Borrielli16}, \cite{Serra18} and a rectangular-shaped geometry endowed with a phononic bandgap structure 
\cite{Tsaturyan17}, both designed and produced in our microfabrication facility. 
The former is produced by bulk/surface micromachining using Deep-RIE and its quality factor is 
mainly limited by the edge losses, while the latter is produced by  wet anisotropic KOH etching and is limited
by distributed losses.  Both membranes have  an intrisic stress of the order of 1 GPa, a dilution factor in the 
range $ \lambda \simeq 10^{-4}$, and features a number of low loss mechanical  modes 
resonating at frequencies in the MHz range with a modal effective mass in the range of pg to ng.    

%-----------------------------------------------------------------------------
\subsection{\label{sec:Chip-scale1} Round-shaped membrane with on-chip shield}
\begin{figure}
\resizebox{0.45\textwidth}{!}{%
  \includegraphics{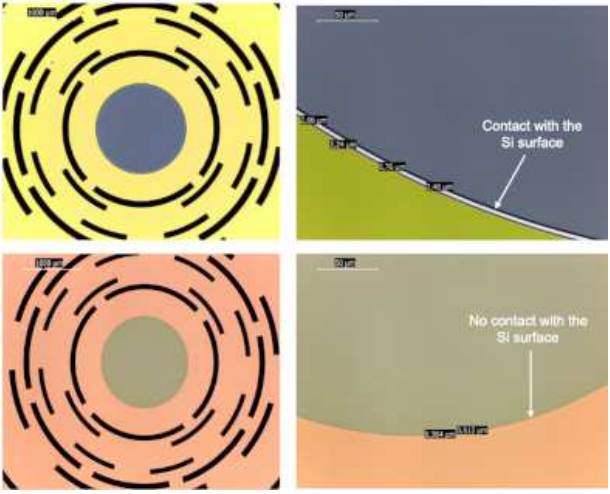}}
\caption{Round-shaped $ Si_3N_4$ membrane surrounded by the on-chip filter and detailed view 
of the membrane's edge. (Top) case of $ Si_3N_4$ deposited on a LPCVD TEOS oxide sacrificial layer 
where the membrane is stitched on the Si subtrate. (Bottom) case of $ Si_3N_4$ deposited on a 
thermally grown oxide without membrane's stitching to the substrate.}
\label{fig:fig6}       
\end{figure}

\begin{figure}
\resizebox{0.45\textwidth}{!}{%
\includegraphics{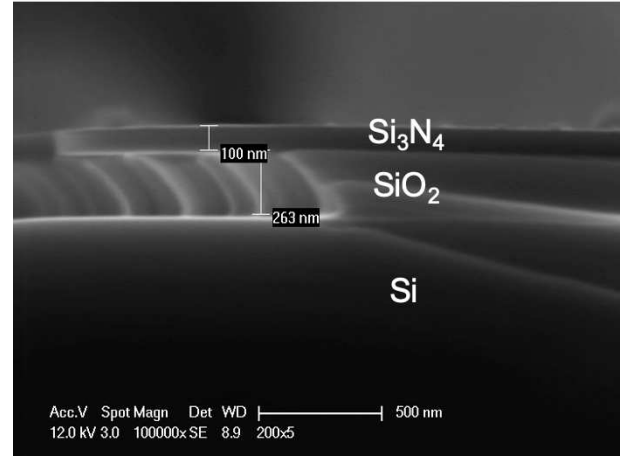}}
\caption{SEM image at the edge of the membrane after partial breakage of a circular membrane showing the undercut of the 
oxide supporting the thin $ Si_3N_4$ membrane.}
\label{fig:fig7}       
\end{figure}

%-----------------------------------------------------------------------------
Tensioned round-shaped membranes were produced with a uniform stress  $\sigma_0 = 0.918 \pm 0.02$ GPa. 
The filtering structure shown in Fig.~\ref{fig:fig6} was patterned on a $ 1 \pm 0.025$ mm thick Silicon-On-Insulator 
(SOI) wafer with Deep-RIE silicon etching Bosch process. We exploit the  2 $ \mu$m thick buried oxide as etch stop layer 
when patterning the spring structure on the front side and the masses in the back side of the wafer. Additional metallic and 
dielectric sacrificial layers were used during processing to protect devices from scratching and from chemicals. 
All structures are patterned using contact aligner lithography and hard-contact exposure in the back side of the wafer. 
The membrane radius is defined by the re-entrant sidewall slopes of successive Deep-RIE steps that enlarges the central hole 
patterned using lithography. Depending on the details of the etching procedure, 
diameter in the range $1500 \pm 150 \,\mu$m can be obtained. In order  to evaluate possible effects of the edge dissipation 
on Q, we fabricated devices with two clamping conditions: membrane directly pinned to the silicon substrate (stitched membranes) 
and membrane without conctact with the substrate. 
 To this aim two oxides were used as sacrificial layers for the Deep-RIE last etching step: 
the LPCVD TEOS oxide shown in Fig.~\ref{fig:fig6} (top) and the thermally grown oxide shown in Fig.~\ref{fig:fig6} (bottom). 
In both cases the sacrificial oxide layer underneath the membrane was removed with an HF aqueous solutions and then rinsed with 
deionised water. Stiching is obtained by undercutting the LPCVD buried oxide below the edges of the window as demostrated by 
Gopalakrishnan.\cite{Gopalakrishnan13}
%------------
%------------
Fig.~\ref{fig:fig7} shows, in case of non-stitched membrane, the initial undercut produced by HF etching of the oxide layer between 
the membrane and the substrate. The cross-section of the profile is irregular and wave-like but at a scale of hundreds of nanometers 
with respect to mm scale of membrane's radius. Given the relatively large etching rate of  LPCVD oxide, about three times greater 
than the thermally grown oxide, it is possible to obtain undercuts extending for $0.6\,-\,3\  \mu$m underneath the membrane. 
Membrane's stitching happens when the capillary forces of liquid trapped in the undercut region overcome the bending rigidity of the membrane. 
The undercut depth is observed to grow approximately linearly with etch time and we found that the minimum length to promote stitching 
is 0.6 $ \mu$m. Smoothness of the contact surfaces  influences the edge losses as well as the presence of polymer residues from Deep-RIE 
process. To ensure repeatability over a number of devices and reduce edge loss, we made aggressive oxygen plasma steps after the last Deep-RIE 
step and flux DI water onto the membrane's edge after the HF release step. 
%------------
%------------
%-------------------------------------------------------------
\subsection{Rectangular-shaped membrane with phononic bandgap} 
\begin{figure}
\resizebox{0.45\textwidth}{!}{\includegraphics{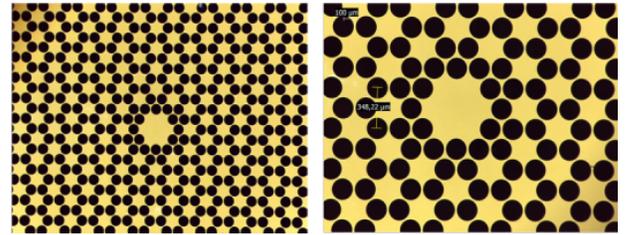}}
\caption{Phononic bandgap lattice (a=346 $ \mu$m) (left) with a detailed view of the central resonating part 
(right) with the hexagonal-shaped defect region in the centre of the lattice.}
\label{fig:fig8}       
\end{figure}

%-------------------------------------------------------------
The device was fabricated following the process described by Tsaturyan,\cite{Tsaturyan17} but we included two extra sacrificial 
layers to increase the process yield. The resulting membrane is 100 nm thick and is supported by a $ 14 \times 14  \ {\rm mm}^2$ 
silicon frame. Details of the structure are shown in Fig. \ref{fig:fig8}. On the same wafer we made devices with different 
lattice constant $ a=(160,\, 346,\, 360,\, 380) \ \mu$m, with corresponding overall membrane size $ 19 \, a \times 19.5 \, a$.

Devices were fabricated on a double-side polished $ 0.525 \pm 0.015$ mm thick wafer with roughness of 0.5 nm RMS. The tensioned 
LPCVD stoichiometric $ Si_3 N_4$ layer, with stress tuned to $ 0.918 \pm 0.02$ GPa, was covered with $4 \mu m$ resist to pattern 
the phononic bandgap lattice by UV lithography in hard contact mode. The silicon nitride layer is then etched with fluorine based plasma. 
Note that the $ Si_3 N_4$ layer is grown directly on the silicon surface without an oxide sacrificial layer as in case of round-shaped device.
The bulk micromachining was done by KOH etching. %using a 35$\%$ KOH at 80$^{\circ}$C 
% to get a reasonable and constant etching rate, that is critical for an optimal release of the membranes. 
and the sacrificial layers on the front side were removed by dry plasma and wet BHF etching. 
With hot DI water we remove residues from the KOH bath and clean the membrane surface. 
We point out that scratches or resist imperfections before UV lithography can trigger the membrane failure during 
the release phase, handling or dicing. 
%------------
%------------

%-----------------------------------------------------------
\section{Q-factor measurement with an interferometric set-up}
\begin{figure}
\resizebox{0.45\textwidth}{!}{\includegraphics{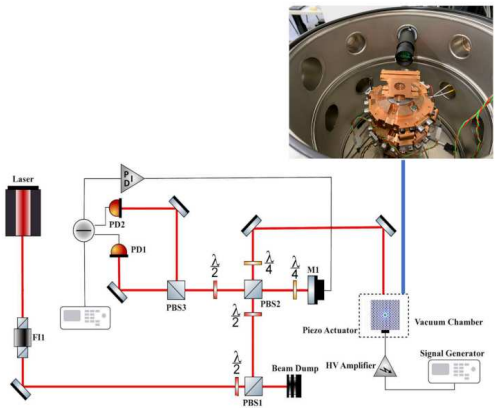}}
\caption{High-sensitivity interferometer scheme for Q-factor measurements. 
(Top) Detailed view of the OFHC copper support for the resonators with the filtering suspension system 
clamped at the bottom of the vacuum chamber.} 
\label{fig:fig9}       
\end{figure}

%-----------------------------------------------------------
The experimental set-up shown in Fig. \ref{fig:fig9} was designed to sense displacements at the thermal noise level for $ \mu \rm g$ to mg-scale devices. 
It consists of a polarisation-sensitive Michelson interferometer followed by a balanced homodyne polarising detection. In details, following the scheme shown in Fig.~\ref{fig:fig9}, a polarising beam-splitter (PBS2) divides the beam into two parts, orthogonally polarised, forming the Michelson interferometer arms. 
The reference arm is focused to an electromagnetically-driven mirror $ M_1$, that is used for phase-locking the interferometer in the condition of maximum displacement sensitivity. 
The sensing arm is instead focused with a waist below 100 $ \mu$m on the oscillator kept in a vacuum chamber at a pressure of $ 10^{-6}$ mbar.  
After the reflection and a double pass through a quarter-wave plate, both beams suffer a 90$^{\circ}$ rotation of their polarisation, 
then they are completely transmitted (reference beam) or reflected (sensing beam) by the PBS2.  
The overlapped beams are then monitored by a homodyne detection, consisting of a half-wave plate, rotating the polarisations by 45$^{\circ}$, 
and a polarising beam-splitter (PBS3) that divides the radiation into two equal parts sent to the photodiodes PD1 and PD2, whose outputs are subtracted. 
The signal obtained is a null-average, sinusoidal function of the path difference in the interferometer. Such a scheme is barely sensitive to laser power fluctuations. 
The difference signal is used as error signal in the locking servo-loop (the locking bandwidth is about 300 Hz) and also sent to the acquisition and measurement instruments. 
In order to measure the Q-factor we first drive the system at the resonance frequency $ \nu_m$ by a piezoelectric actuator mounted on the sample holder, then remove the drive and acquire the amplitude of the displacement signal using the interferometer. The mechanical vibration follows an exponentially 
damped decay whose envelope amplitude varies according to: 
$ u(t)=u_0 \exp ({-t/t_m})$ where $ t_m=Q/(\pi \nu_m)$ is the decay time of the mode. 
%------------
%------------
%---------------------------
\section{Membrane inside an optical cavity}
\begin{figure}
\resizebox{0.45\textwidth}{!}{\includegraphics{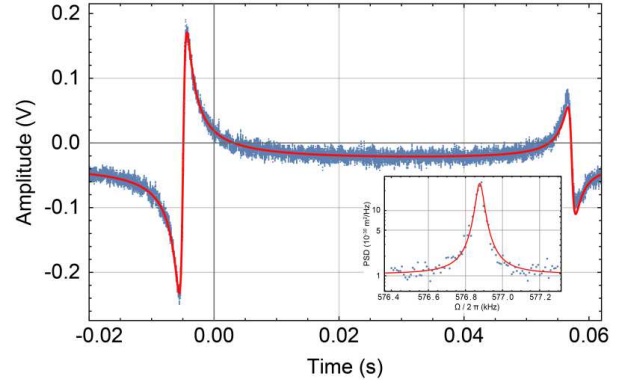}}
\caption{Pound-Drever-Hall signal from an optical cavity containing a membrane with phononic bandgap structure. 
The distance between the carrier (right dispersive signal) and sideband (left signal) is 13.3 MHz, and is used 
to calibrate the frequency scan. Inset: power spectral density (PSD) of the A mode.} 
\label{fig:fig10}       
\end{figure}

%---------------------------
The use of membrane oscillators in high sensitivity optomechanical nanosensors implies that they should be located inside 
high Finesse optical cavities, as partially reflecting surfaces. As a consequence, they should not spoil the cavity optical quality, 
and their overall mechanical noise (including the sensing mode and all the other mechanical modes) should be low enough to allow 
a stable frequency locking between the cavity optical resonance and the laser field. In order to test these requirements we have 
mounted the dices incorporating the membranes inside plano-concave cavities, using silicon spacers between the membrane dice an the 
flat mirror to guarantee parallelism and mechanical stability. The round membranes with acoustic shield have been successfully used 
even inside helium flux cryostats, in quantum optomechanics experiments.\cite{Delic20, Chowdhury19, Vezio20, Bonaldi20} Concerning the rectangular 
membranes with phononic bandgap structure, we have tested the PM-360 sample inside a 48 mm long cavity, in high vacuum at room temperature. We show in Fig. \ref{fig:fig10} the Pound-Drever-Hall signal obtained in this setup, by phase modulating the probe laser at 13.3 MHz. The fit to the data gives a cavity linewidth of 250 kHz (Finesse 12000), setting the bandgap modes in the resolved sidebands regime, with optical losses dominated by the input mirror transmission. The laser was then  stably locked to the cavity. In the inset of the Fig. \ref{fig:fig10} we show a portion of the spectrum of 
the same Pound-Drever-Hall signal, exhibiting the resonance of the A mode, slightly optically cooled by an additional weak, red-detuned beam.  
%------------
%------------
%------------
%---------------------------
\section{RESULTS AND DISCUSSION}
%---------------------------
\label{sec:RD}
In this section we compare the performances of the two optomechanical sensors in terms of $ Q \times \nu$ product 
and force sensitivity. We discuss advantages and drawbacks also in terms of modal density, surface's functionalization and microfabbrication process complexity. 
%----------------------------------------------
\subsection{Figure of merit $ Q \times \nu$}  
\begin{figure}
\resizebox{0.50\textwidth}{!}{\includegraphics{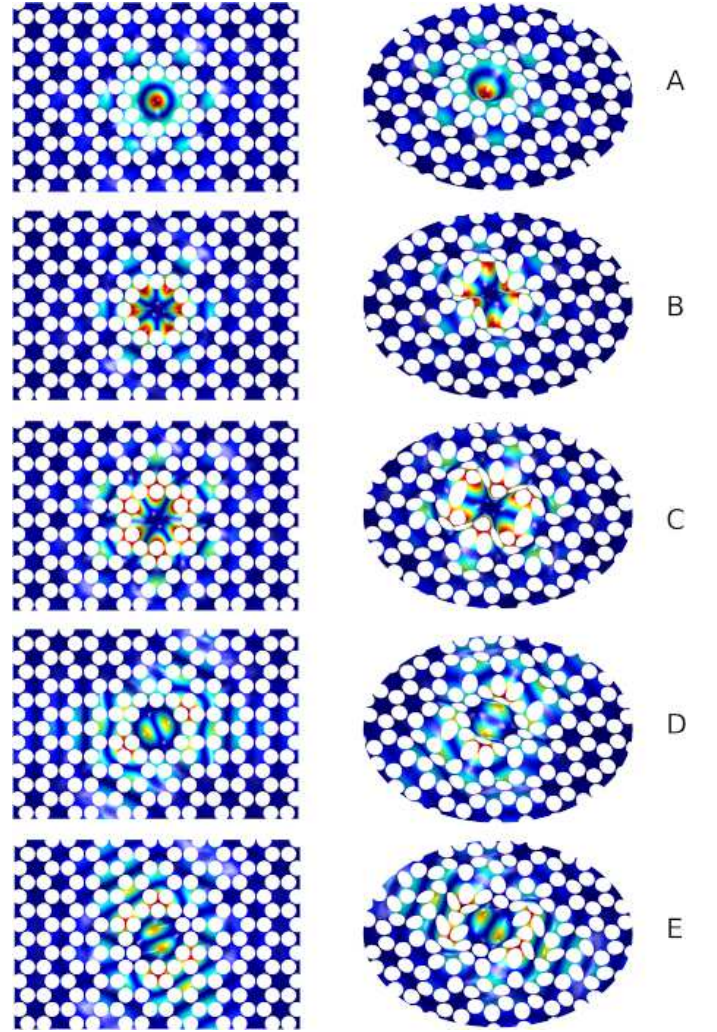}}
\caption{FE modal analysis with $ a=160 \ \mu$m of bandgap modes with frequencies 
$ \nu_A > \nu_B > \nu_C > \nu_D > \nu_E$ showing that the defect mode evanescently couples 
to substrate at the edge (soft-clamping). Modes B and C and D and E are doublets.}
\label{fig:fig11}       
\end{figure}

%----------------------------------------------
In round-shaped membranes, modal frequencies are calculated by the well-known analytical formula: 
$ \nu_{nm}=\frac{1}{\pi 2 R} \alpha_{nm} \sqrt{\frac{\sigma}{\rho}}$
where $ \alpha_{nm}$ is the root of the Bessel function $ J_m$ and $ (n,m)$ are the radial and circumferential modal indexes. 
For the rectangular membranes with phononic crystal, the modal frequencies are numerically evaluated by Finite Element Analysis (FEA).  
The modal shapes of the five out-of-plane modes in the bandgap are shown in Fig.~\ref{fig:fig11} where the first mode A is a singlet while 
(B, C) and (D, E) are doublets. Similarly to what shown for mode A  in Fig. ~\ref{fig:fig4}, all bandgap modes are well localized in the 
defect region and the displacement of the membrane fades out by going towards the edge, thus achieving the soft-clamping condition.

\begin{figure}
\resizebox{0.45\textwidth}{!}{\includegraphics{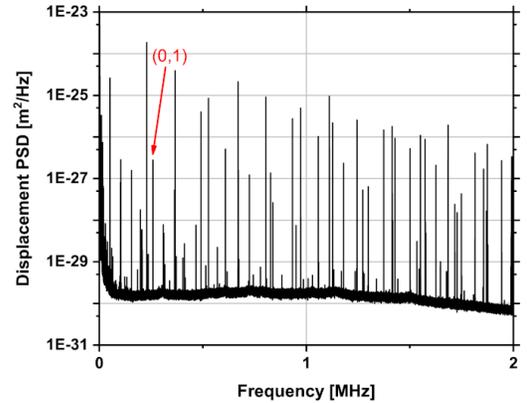}}
\caption{Displacement PSD for the round-shaped membrane. Data are derived from the FFT of the interferometer output.}
\label{fig:fig12}       
\end{figure}

\begin{figure}
\resizebox{0.45\textwidth}{!}{\includegraphics{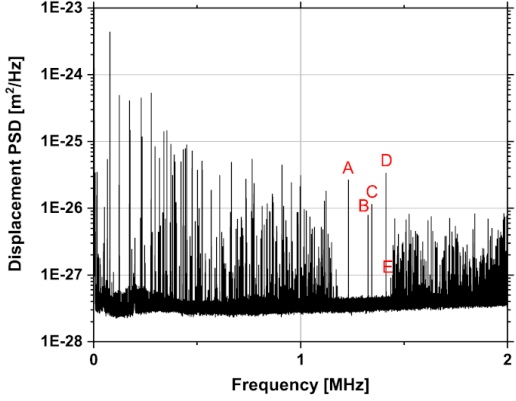}}
\caption{Displacement PSD for a rectangular-shaped membrane with $ a=160\ \mu m$. Data are derived from the FFT of the interferometer output.}
\label{fig:fig13}       
\end{figure}

In Fig.~\ref{fig:fig12} and \ref{fig:fig13} we compare the Power Spectral Density (PSD) of the displacement of the central area of 
the devices, measured for acoustic frequencies up to 2 MHz. Here the modes are driven by the thermal force to an amplidude depending 
upon the modal equivalent mass and upon the coupling with the laser beam. 

In case of the round-shaped membrane, the normal modes are well separated and emerge from the background noise of $ 10^{-30} \ {\rm m}^2/$Hz 
(Fig.~\ref{fig:fig12}). In general almost all expected modes can be found at the expected modal frequency; in the Figure we show for 
instance the position of the first drum mode (n,m)=(0,1). Here the filtering stage effectively  decouples the frame from the membrane, 
as described by Borrielli,\cite{Borrielli16}  hence the clamping losses are made negligible and the membrane dynamics 
is dominated by the intrinsic dissipation, both at the clamping edge and in the central area, for all mechanical modes. 
The Q factor is in the range between $ 5-10 \times 10^6$ and no differences could be appreciated in case of stitched and non-stitched membranes. 

\begin{figure}
\resizebox{0.45\textwidth}{!}{\includegraphics{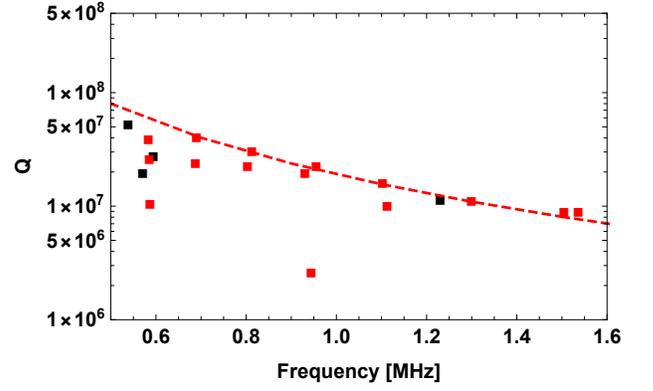}}
\caption{Measured Q factors vs frequency for the bangap mode A of a rectangular-shaped phononic membrane 
of thickness $ h=100$ nm and stress $ \sigma=0.918$ GPa. Red line is the theoretical curve derived using Eq. 
(\ref{eq:Q_Qint_soft-clamping2D}) and sets a theoretical limit for the 2-D phononic membrane.  
Data points are experimental results measured in our set-up (black squares) or measured by Tsaturyan,\cite{Tsaturyan17} (red squares). 
In this case, the experimental values are scaled according to Eq. (\ref{eq:Qf_SCALING}) to allow a proper comparison when the
 membrane thickness or the internal stress is not the same as our devices.}
\label{fig:fig14}    
\end{figure}

The rectangular-shaped membrane PSD spectrum, shown in Fig.~\ref{fig:fig13}, has a bandgap region with some distinct peaks emerging from 
the background noise (here at a level of $ 10^{-27} \ {\rm m}^2/$Hz to $ 10^{-28} \ {\rm m}^2/$Hz), each corresponding to the out-of-plane modes  
(A, B, C, D, E). In this case the Q-factor can be as high as $ 5 \times 10^7$ and scales down quadratically with the lattice constant $ a$.
 In the region outside the bandgap the modal density dramatically increases due to a large number of resonances of the phononic crystal. 
%------------
%------------
For the bandgap modes we observed a dispersion of Q value  in agreement with results obtained by Tsaturyan.\cite{Tsaturyan17} In Table \ref{tab:tab1} we report the results for two rectangular-shaped  membranes with lattice 
constant $a=160$ $ \mu$m and of $a=360$ $ \mu$m. The performances of the devices presented in this work are fully 
consistent with those presented by Tsaturyan,\cite{Tsaturyan17} if the physical parameters of the membranes are 
properly scaled according to the theory presented in Sec. \ref{sec:damping4}:
%------------
\begin{equation}
\label{eq:Qf_SCALING}
 \nu =\hat{\nu}\times \frac{\hat{a}}{a}\sqrt{\frac{\sigma}{\hat{\sigma}}}, \ \ \ Q=\hat{Q} \times \frac{\sigma \hat{h} a^2}{\hat{\sigma} h\hat{a}^2}
\end{equation}
%------------
where the paramenters ($ \nu, Q, h, a, \sigma $)  and ($ \hat{\nu}, \hat{Q}, \hat{h}, \hat{a},\hat{\sigma} $) identify two different geometrical configurations
of the device.  
%------------------
%----------------
\begin{table}[htb]
\caption{Measured Q-factor bandgap modes A, B, C, D, E for square-shaped membranes with lattice constant $ a=160, 360\, \mu$m 
with membrane size $\rm 3040 \times 3120 \,\mu {\rm m}^2$ and $\rm 6840 \times 7020\, \mu {\rm m}^2$ respectively.} 
\label{tab:tab1}
\begin{center} 
\begin{ruledtabular}    
\begin{tabular}{ccccccc}
\rule[-1ex]{0pt}{3.5ex}  MODE & \multicolumn{3}{c}{PM-160} & \multicolumn{3}{c}{PM-360}\\
\hline
\rule[-1ex]{0pt}{3.5ex}       & $ \rm \nu $ & $ Q$ &  $ \nu \times Q $  & $ \nu $ & $ Q$ &  $ \nu \times Q $\\
\rule[-1ex]{0pt}{3.5ex}       & $\rm [MHz]$ & $ 10^6$ &  $\rm [THz]$  & $\rm [MHz]$ & $ 10^6$ &  $\rm  [THz]$\\
\hline
\rule[-1ex]{0pt}{3.5ex}  A & 1.23 &  11.2  & 11.19 & 0.571 & 19.3 & 11.02\\
\rule[-1ex]{0pt}{3.5ex}  B & 1.32 & 10.4 & 13.73  & 0.614 & 11.0 & 6.75\\
\rule[-1ex]{0pt}{3.5ex}  C & 1.34 &  11.5 & 15.41 & 0.620 & 54.0 & 33.48 \\
\rule[-1ex]{0pt}{3.5ex}  D & 1.41 & 3.2 & 4.48 & 0.649 & 10.3 & 6.68\\
\rule[-1ex]{0pt}{3.5ex}  E & 1.43 &  11.4 & 16.30 & -- &  -- & --\\
\end{tabular}
\end{ruledtabular}
\end{center}
\end{table}
%------------------
Some interesting considerations can be drawn by comparing the estimates obtained from the models developed 
in section \ref{sec:damping4} with the measurements reported in Table \ref{tab:tab1}. 
In Fig.~\ref{fig:fig14} we show the Q-factor measured for the mode A of a membrane with $ \sigma=0.918 \  $GPa, 
thickness $ h=100\ $nm and lattice parameter $ a=360 \,\mu$m, together with properly scaled values for similar devices.\cite{Tsaturyan17}
We also show the  estimate obtained for mode A  from the 2-D model 
Eqs (\ref{eq:Q_Qint_soft-clamping2D}) with $\phi=2.66 \times 10^{-4}$ and $\beta=60 \times 10^9 \, {\rm m}^{-1} $.
%------------
%------------
%------------
%------------ 
In Fig.~\ref{fig:fig14} we compare the two devices in terms of the figure of merit  $ Q \times \nu$, that measures 
the coherence rate. Devices with $Q \times \nu$ lying inside the gray area are compliant with the requirement for room temperature 
optomechanics,  $ Q \times \nu >$  6.2 THz.\cite{Aspelmeyer14} Almost all modes of round-shaped membranes met the requirement, 
apart from some mode at low frequency. Bandgap modes of the rectangular-shaped membrane,  where the quality factor is limited by the 
distributed loss mechanism, are fully compliant but more scattered results between the different modes, at least for what observed 
with the devices obtained with our production process. We point out that the quality factor can be tuned by changing the overall 
dimension of the device, as in larger membranes the edge loss becomes relatively smaller. On the other hand, we believe that this 
comparison is fair because we produced, for both kind of devices, the larger structures that could be hosted on our standard 
silicon frame 14x14 mm$^2$.

%-----------------------------
\subsection{Force sensitivity} 
%----------------------------- 
The opto-mechanical membrane is an elastic body that vibrates due to the combined effects of the 
external forces, back-action radiation pressure and the thermal Langevin force.  
When the system is used as a probe for an external force, its sensitivity is limited by the equivalent input
force density due to the technical noises. Given that we compare devices of different geometry and size, 
the evaluation must account for the precise shape of the surface 
sampled by the readout and for the point of application of the external force. 
Here we neglect readout backaction noise and consider only a force orthogonal to the membrane's plane and with a 
frequency equal to one of the membrane's resonances, say $ \omega_n$, but even the most general case could be evaluated, 
as done for instance in the context of resonant detectors of gravitational waves.\cite{Bonaldi06}
%----------------------------
\subsubsection{Generic force}
%----------------------------
Starting from the normal mode expansion (\ref{eq:espansione}),  we evaluate the displacement of the membrane when a volume force density  $\textbf{G}(\textbf{r})$ is applied to the membrane. The total driving force comes from the integration of this force density over the volume of the membrane $V$. If the time evolution of the force can be separated, we write:
\begin{equation}
\label{eq:forcefactorize}
{\bf G}({\bf r},\,t)=G_t(t)\;{G_r}({r},\,\theta,\,z) \,\textbf{i}_z
\end{equation}
where $G_t(t)$ is the total force applied to the membrane and  $G_r$ is a density function describing how the force field varies in the membrane's volume, with normalization:
\begin{equation}
\label{eq:gnorm}
\int_V dV {G_r}({r},\,\theta,z)=1
\end{equation}
Thanks to the separability of time from spatial variables, the time evolution of the coefficient 
$a_{n}(t)$  is equivalent to the time development of a forced harmonic oscillator:
\begin{equation}
\label{eq:timedep}
M \,\frac{\partial^2 a_n(t)}{\partial t^2}+M\,\omega_n^2\,
a_n(t)\,
=\,G_t(t)\,  \int_V\,\textit{d}V \,{G_r}( {r},\,\theta,\,z)\,  {w}_n({r},\,\theta)\,.
\end{equation}
where ${w}_n({r},\,\theta)$ is constant along the membrane's thickness. 
Note that, in order not to burden the notation, we identify the mode by a single index $n$ in this section.
In the frequency domain (here and in the following we indicate the Fourier transform with a tilda) we have then:
\begin{equation}
\label{eq:qsolution}
\widetilde{a}_n(\omega)\,=\,\frac{1}{M}\frac{\widetilde{G}_t(\omega)}{(\omega^2_{n}-\omega^2)+
i \omega^2_{n}
\phi_{n}(\omega)}\,   \int_V \textit{d}V\,{G}_r({r},\,\theta,\,z)\;{w}_n({r},\,\theta)
\end{equation} 
where the system loss  is modeled in the frequency domain by including explicitly the damping term $\phi_n=1/Q_n$.\cite{Saulson90}
If we consider the  output coordinate $X$  defined in Eq. (\ref{eq:XresponseV}), we see that, at the resonance $\omega_m$,  the response $T_G(\omega)=\widetilde{X}/\widetilde{G}_t$ of the oscillator to the external force density $G$ is:
%----------------
\begin{eqnarray}
 T_G(\omega_n)=&& -i \, \frac{Q_n}{M \omega_n^2}   \int_V \textit{d}V\,{G}_r({r},\,\theta,\,z)\;{w}_n({r},\,\theta) \nonumber \\
&& \times \int_S dS\,  {P}_s({r},\,\theta) \,{w}_n({r},\,\theta)
\label{eq:eqTn2}
\end{eqnarray}
%---------------- 
where $ {P}_s({r},\,\theta)$ is the weight function of the readout,  defined in Eq. (\ref{eq:G}) for an optical readout.
%--------------------------------------------
\subsubsection{Thermal noise and sensitivity}
%--------------------------------------------
The fluctation-dissipation theorem states that the one-side noise PSD of the coordinate $X$ at the angular frequency $ \omega$ is:   
%---------------
\begin{equation}
 S_{XX}(\omega)= -\frac{4 k_B T}{\omega} \Im\{T_F(\omega)\}
\label{eq:PSD}
\end{equation}
%---------------
where $ T_F(\omega)=\widetilde{X}(\omega)/\widetilde{F}_t(\omega)$ is the transfer function between the 
readout observable $X$ and a force $F$ applied with the same weight function ${P}_s$ of the readout:\cite{levin}
\begin{equation}
{{F}}({r},\,\theta,\,t)=F_t(t)\, {P}_s({r},\,\theta)  
\end{equation}
If we substitute this specific force field in the general solution Eq. (\ref{eq:eqTn2}), we find, at resonance
 $ \omega_n$: 
%----------------
\begin{equation}
 T_{F}(\omega_n)= -i \frac{Q_n}{M \omega_n^2} \left( \int_S dS\,  {P}_s({r},\,\theta) \,{w}_n({r},\,\theta)\right)^2=-i\frac{Q_n}{\omega_n^2 m^\textrm{eff}_n} \\
\label{eq:eqTnTh}
\end{equation}
%---------------- 
where the effective mass has been defined in Eq. (\ref{effmass}).
From Eqs. (\ref{eq:PSD}) and (\ref{eq:eqTnTh}), the  
thermal noise PSD on the output variable
is: 
%---------------
\begin{equation}
 S_{XX}(\omega_n)= 4  k_B T \frac{Q_n}{m_n^{\rm eff}\, \omega_n^3}
\label{sxx_on}
\end{equation}
%---------------
where $m_n^{\rm eff}$  for the circular membrane and for the gap modes of the 
the rectangular soft-clamped membrane are evaluated respectively in Eqs. (\ref{modalMassCircle}) and (\ref{modalMassA}).
The minimal detectable force is obtained by the comparison between the ouput $X$ induced by the force  $\textbf{G}$  and the noise power spectral density $S_{XX}$. The sensitivity of the device as a detector of this force is defined as the ratio:
\begin{equation}
S_{GG}(\omega_n)=\frac{S_{XX}(\omega)}{|T_G(\omega)|^2}
\end{equation}
and should be as low as possible in the frequency range of interest.
At resonance $\omega_n$ we can substitute equations (\ref{eq:eqTnTh}) and (\ref{sxx_on}) and obtain:
\begin{equation}
S_{GG}(\omega)=4  k_B T  \frac{\omega_n}{Q_n}\,\,\frac{M}{\left(\int_V \textit{d}V\;{G}_r({r},\,\theta,\,z)\;{w}_n({r},\,\theta)  \right)^2}
\label{eq:SGG}
\end{equation}
that make clear the role of the modal shape on the sensitivity of the system. It is not surprising that low temperatures and high quality factors are of benefit to the measure, while the term depending on the modal shape is sometimes overlooked. 
%--------------------------
\subsubsection{Point force}
%--------------------------
To study the case of  a force  applied in the center of the membrane, we assume for simplicity that it is directed along the membrane axis $\textbf{i}_z$ and applied onto the surface $z=0$ with the shape of the readout weight function $P_s(r,\,\theta)$. The correspondent force density is $\textbf{U}(\textbf{r},t)= U_t(t) P_s(r,\theta)\, \delta(z)\,\textbf{i}_z$, where $U(t)$ is the total applied force. The normalization of the density function is a consequence of the normalization of the weight function $P_s$. In this case equation \ref{eq:SGG} can be rewritten as:
%---------------
\begin{eqnarray}
S_{UU}(\omega_n)=&& \, 4  k_B T  \frac{\omega_n}{Q_n}\,\,\frac{M}{\left(\int_S dS\,  {P}_s({r},\,\theta) \,{w}_n({r},\,\theta)\right)^2} \nonumber \\
=&& \, 4  k_B T  \frac{\omega_n}{Q_n}m^{\rm eff}_n
\label{eq:SUU}
\end{eqnarray}
%------------
The sensitivity of both devices is reported in Table \ref{tab:tab2}, where the
results have also been extrapolated to cryogenic (T = 4.2) and ultracryogenic (T=14 mK) temperatures. 
%-------------
\begin{table*}
\caption{Thermal noise limits to the sensitivity to a point force at room, cryogenic and ultra-cryogenic temperatures. RS-M: mode (0,1) of round shaped membrane resonator. PM-$a$: mode A of soft-clamped membrane with lattice parameter $a$. The laser waist $w_0$ of the read-out and the effective mass of the oscillator are reported in the table.
The physical parameters of the devices are listed here. RS-M: \ $ h=100\, {\rm nm},\, f=284\,{\rm  kHz},\, Q=8.7 \times 10^6$;  
PM-160: $ h=100\, {\rm nm},\, f=1.23\, {\rm MHz},\, Q=9.1 \times 10^6$; PM-346: $ h=100\,  {\rm nm},\, f=630\, {\rm kHz}, \,Q=19.3 \times10^6$.}
\label{tab:tab2} 
\begin{ruledtabular}
\begin{tabular}{lccccc}
\rule[-1ex]{0pt}{3.5ex} $ S_{UU}^{1/2} $& & & \textcircled{a} 300 [K] & \textcircled{a} 4.2 [K] & \textcircled{a} 14 [mK] \\
\rule[-1ex]{0pt}{3.5ex}  & $ w_0 [\mu m]$ & $ m_{eff} [ng] $ &  $  [aN/\sqrt{Hz}] $& $  [aN/\sqrt{Hz}] $ & $ [aN/\sqrt{Hz}] $\\
\hline
\rule[-1ex]{0pt}{3.5ex} RS-M & 600 & 326 & 1014.13 & 119.98 & 6.92\\
\rule[-1ex]{0pt}{3.5ex} RS-M & 150 & 132 & 645.42 & 76.37 & 4.41 \\
\rule[-1ex]{0pt}{3.5ex} RS-M & 50 & 125 & 627.85 & 74.29 & 4.28 \\
\hline
\rule[-1ex]{0pt}{3.5ex}  PM-160 & 50 & 14.7 & 438.41 & 32.81  & 0.96\\
\rule[-1ex]{0pt}{3.5ex}  PM-346 & 50 & 38.0 & 329.51  & 24.61  & 0.72\\
\end{tabular}
\end{ruledtabular}
\end{table*}
%------------------
We have extrapolated  from literature data  the temperature dependence of the Q-factor for the soft-clamped membranes.\cite{Tsaturyan17} On the contrary we have assumed a   constant value of Q for round-shaped membranes.\cite{Borrielli16} Soft clamped membranes are in general more sensitive, thanks to their higher quality factor, but in both cases these figures outperform the pN sensitivities typical of atomic force microscopes, and push us to further improve the performance and stability of the production process. 
%------------
%------------
\begin{figure}
\resizebox{0.45\textwidth}{!}{\includegraphics{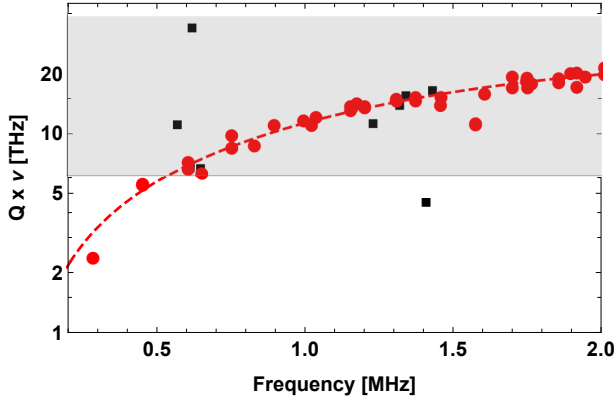}} 
\caption{The plot shows $ Q \times \nu$ vs. frequency for the two membranes, at room temperature.  
Red circles refers to round-shaped membrane, black squares refers to modes (A, B, C, D)
of soft-clamped membrane with lattice $ a=160\, \mu {\rm m}$ and $ a=360\, \mu {\rm m}$. 
The red-dashed line is the fitting model for the round shaped membranes. 
Data points within the gray shaded area fulfill the minimum requirement for quantum optomechanics
at room temperature.}
\label{fig:fig15}    
\end{figure}
%---------------------------------
\subsubsection{Uniform force field}
%---------------------------------
We consider now the sensitivity of the membranes to a uniform force density field oriented along the membrane axis,
 $\textbf{i}_z$. 
%------------
\begin{figure}
\resizebox{0.45\textwidth}{!}{\includegraphics{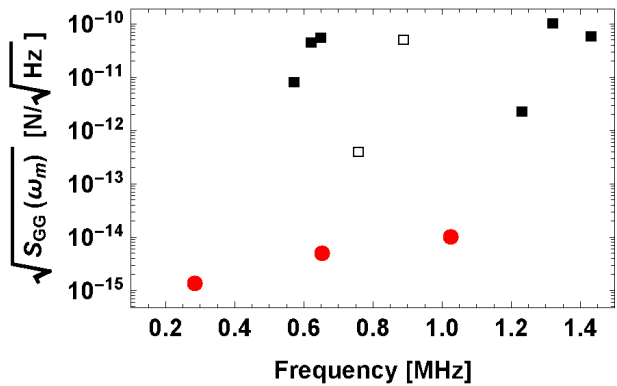}}
\caption{Calculated force sensitivity $S_{GG}(\omega_m)$ at a temperature of 300 K for membrane based devices. 
Red circles correspond to lowest frequency  axisymmetric modes of the round-shaped membrane while black squares correspond to bandgap modes A, B, C, D, E of the soft-clamped membrane. The empty black square refers 
to soft-clamped membrane with $ a=346 \ \mu {\rm m}$ and thickness $ 35 \ {\rm nm}$ described by Tsaturyan.\cite{Tsaturyan17}}
\label{fig:fig16}    
\end{figure}
%------------
In this case we have  $G_r(r,\,\theta,\,z)=1/V$ and  Equation  (\ref{eq:SGG}) becomes
\begin{equation}
S_{GG}(\omega)=4  k_B T  \frac{\omega_n}{Q_n}\,\,\frac{M \,V^2}{\left(\int_V \textit{d}V\;{w}_n({r},\,\theta)  \right)^2}
\label{eq:SGGuniform}
\end{equation}
%------------
that can be evaluated by FEM for the relevant modes. Here we focus on the vibrational modes 
in the bandgap of the soft-clamped membrane and on the axisymmetric modes $(0,\,n)$ of the round-shaped membrane.  Results are summarized in Figure (\ref{fig:fig16}), where we show the predictions for the membranes we have produced the laboratory and the ultra-high Q soft-clamped membrane described by Tsaturyan,\cite{Tsaturyan17} featuring  Q factors well over $10^7$.
In general round-shaped membranes can detect an external uniform force 
with much higher sensitivity with respect to bandgap modes of the soft-clamped membrane. This is due to the   the peculiarity
of soft-clamped membranes, where the matching of the defect modes and the phonon modes produce a modal shape that is confined in the defect area, as it is shown in Figure (\ref{fig:fig5}) for bandgap mode A.
 On the contrary, the axysimmetric modal shapes $(0,n)$  of round-shaped membranes are distributed over the whole membrane and allow a better matching of the modal shape with the uniform force field, even though their sensitivity decreases
with the number of circumferential nodes.   
 An estimation of the S/N in the 
case of specific force signals goes beyond the scope of the article.

%--------------------
\section{Conclusions}
%--------------------
Freestanding nano-membranes initially became widespread as windows for chemical analysis by x-ray, e-beam and TEM, 
but recent studies have shown how useful they can be as highly coherent mechanical resonator  in the field of quantum 
technologies and fundamental physics. In these optomechanical systems the  membrane is embbeded in a Fabry-Perot optical 
cavity, where the radiation pressure couples the membrane's mechanical displacement with  the intracavity field.
The high value of the mechanical quality factor, at cryogenic and room temperatures, guarantees basic requirements for many 
challeging experiments, such as ground state sideband cooling, quantum non demolition measurements and quantum squeezing of 
resonator states. Recent efforts are directed toward the functionalization of SiN membranes, to allow the coupling with 
external signals. We mention for instance the deposition of a metal layer for the realization of device capables of converting 
RF/microwaves to optical signals, possibly at the quantum limit.\cite{Moaddel18}

In this work we investigated the mechanical dissipation of two state-of-the-art membrane-based devices used in quantum 
optomechanical experiments, both produced in our microfabrication facility. They are based on tensioned SiN membranes, with 
equal thickness and stress, and reach high  Q-factors thanks to the dissipation dilution effect.\cite{Saulson90} Concepts 
based on stress/strain engineering and soft-clamping were used in the design to preserve this feature from the influence of 
the environment. A general design framework is outlined and discussed, showing advantages and drawbacks. We described the 
fabrication processes based on bulk/surface silicon micromachining and discussed our experimental results in the context of 
literature data.

Resonators with soft-clamping ensure the achievements of very high Q factor for a number of modes inside the bandgap, on the other 
hand we have observed that they are more fragile, as imperfections in holes edge can trigger the membrane failure during the release 
phase, handling or dicing. Currently, in the outcomes of our production process, the soft-clamped membranes are more sensitive to 
fabrication issues and we are reviewing the process to increase reproducibility and yield.

Circular-shaped membrane with loss shield constitutes a reliable platform for optomechanical experiments, 
with modal Q-factor almost constant over a large frequency band, but did not demonstrated quality factor higher than $10^7$. 
In this case  we are studying how to apply a stress engineering approach to 
increase the quality factor for a membrane while maintaining a membrane thickness of 100 nm in the central region, as it is needed 
to facilitate the coupling with the laser beam.

We have completed our comparison with the evaluation of the interaction with an external force.  As usual, we define the force 
sensitivity as the ratio between the thermal noise PSD and the square of the response function to the force. In the calculation 
we neglect back-action noise from the readout.
The analysis show that the sensivitity is strongly dependent on the design of a device, with soft clamped membranes better suited 
for the detection of point forces while round shaped membranes have the best performance in the detection of uniform force fields. 
This issue is relevant in proposals aiming to use low-loss resonators in the detection dark matter signatures,\cite{carney21} with an approach similar to that  deleloped for a massive antenna.\cite{aurigaDM} We also note that 
in round-shaped membranes the Q-factor is obtained for a large number of modes, providing an ideal platform for multimode quantum 
optomechanics or multimode sensing.\cite{Moaddel18}

\begin{acknowledgments}
Research was performed within the Project QuaSeRT funded by the QuantERA ERA-NET Cofund in Quantum Technologies
implemented within the European Union’s Horizon 2020 Programme. The research has been partially supported by INFN
(HUMOR project).
\end{acknowledgments}

%-------------------

%-------------------
\end{document}